\documentclass[twocolumn,twoside,slac_two]{revtex4}
\usepackage{graphicx}
\usepackage{fancyhdr}
\newcommand{\Dzb}{\overline{D^0}}
\newcommand{\barD}{\overline{D^0}}
\newcommand{\DzDzb}{D^0-\overline{D^0}}

\newcommand{\bea}{\begin{eqnarray}}
\newcommand{\eea}{\end{eqnarray}}
\newcommand{\beq}{\begin{equation}}
\newcommand{\eeq}{\end{equation}}

\begin{document}

\title{CP violation in charm}

%

\author{Alexey A Petrov}
\affiliation{Department of Physics and Astronomy, Wayne State University, Detroit, MI 48201, USA}
%
%
\begin{abstract}
CP-violating asymmetries in charm provide a unique probe of physics
beyond the Standard Model. I review several topics relevant to searches 
for CP-violation in charmed meson and baryon transitions. 

\end{abstract}

\maketitle

\thispagestyle{fancy}


\section{Introduction}

Charm transitions play a unique dual role in the modern investigations of flavor physics.
They provide valuable supporting measurements for studies of CP-violation in $B$-decays, 
such as formfactors and decays constants, as well as outstanding opportunities for 
indirect searches for physics beyond the Standard Model (SM). 
It must be noted that in many dynamical models of new physics the effects of new 
particles observed in $s$, $c$, and $b$ transitions are correlated. Therefore, such 
combined studies could yield the most stringent constraints on parameters of those models. 
For example, loop-dominated processes such as $\DzDzb$ mixing or flavor-changing 
neutral current (FCNC) decays are influenced by the dynamical effects of 
{\it down-type particles}, whereas up-type particles are responsible 
for FCNC in the beauty and strange systems. Finally, from the practical point of 
view, charm physics experiments provide outstanding opportunities for studies of 
New Physics because of the availability of large statistical samples of data.

CP-violation can be introduced in Quantum Field Theory in a variety of 
ways~\cite{BigiSandaBook}. 
One way, CP-violation can be introduced explicitly through dimension-4 
operators (the so-called ``hard'' CP-breaking). This is how CP-invariance is 
broken in the Standard Model via quark Yukawa interactions,
\beq
{\cal L}_Y = \xi_{ik} \overline \psi_i \psi_k \phi + \mbox{~h.c.}
\eeq
The complex Yukawa couplings $\xi_{ik}$ lead to complex-valued 
Cabibbo-Kobayashi-Maskawa (CKM) quark mixing matrix providing the
natural source of CP-violation for the case of the Standard Model with three 
(or more) generations. Another way could be via operators of dimensions less than 
four (the ``soft'' CP-breaking), which is popular in supersymmetric models.
Yet another way is to break CP-invariance spontaneously. This method, which 
is somewhat aesthetically appealing, introduces CP-violating ground state with
CP-conserved Lagrangian. It is realized in a class of left-right-symmetric models 
or multi-Higgs models. All these mechanisms can be probed in charm transitions.
In fact, observation of CP-violation in the current round of charm experiments
is arguably one of the cleanest signals of physics beyond the Standard Model (BSM).

It can be easily seen why manifestation of new physics interactions in the charm
system is associated with the observation of (large) CP-violation. This 
is due to the fact that all quarks that build up the hadronic states in weak 
decays of charm mesons belong to the first two generations. Since $2\times2$ 
Cabibbo quark mixing matrix is real, no CP-violation is possible in the
dominant tree-level diagrams which describe the decay amplitudes. 
CP-violating amplitudes can be introduced in the Standard Model by including 
penguin or box operators induced by virtual $b$-quarks. However, their 
contributions are strongly suppressed by the small combination of 
CKM matrix elements $V_{cb}V^*_{ub}$. It is thus widely believed that the 
observation of (large) CP violation in charm decays or mixing would be an 
unambiguous sign for new physics. This fact makes charm decays a valuable 
tool in searching for new physics, since the statistics available in charm 
physics experiment is usually quite large.

As with other flavor physics, CP-violating contributions in charm can be generally 
classified by three different categories:

\begin{enumerate}
\item[(I)]
CP violation in the $\Delta C =1$ decay amplitudes. This type of CP violation 
occurs when the absolute value of the decay amplitude for $D$ to decay to a 
final state $f$ ($A_f$) is different from the one of corresponding 
CP-conjugated amplitude (``direct CP-violation''). This can happen if
the decay amplitude can be broken into at least two parts associated with 
different weak and strong phases,
\beq\label{DirectAmpl}
A_f =
\left|A_1\right| e^{i \delta_1} e^{i \phi_1} +
\left|A_2\right| e^{i \delta_2} e^{i \phi_2},
\eeq
where $\phi_i$ represent weak phases ($\phi_i \to -\phi_i$ under CP-transormation),
and $\delta_i$ represents strong phases ($\delta_i \to \delta_i$ under CP-transformation).
This ensures that CP-conjugated amplitude, $\overline A_{\overline f}$ would differ 
from $A_f$.

\item[(II)] 
CP violation in $\DzDzb$ mixing matrix. Introduction of 
$\Delta C = 2$ transitions, either via SM or NP one-loop or tree-level NP 
amplitudes leads to non-diagonal entries in the $D^0-\barD$ mass matrix,
\beq\label{MixingMatrix}
\left[M - i \frac{\Gamma}{2} \right]_{ij} = 
\left(
\begin{array}{cc}
A & p^2 \\
q^2 & A 
\end{array} 
\right)
\eeq
This type of CP violation is manifest when 
$R_m^2=\left|p/q\right|^2=(2 M_{12}-i \Gamma_{12})/(2 M_{12}^*-i 
\Gamma_{12}^*) \neq 1$.

\item[(III)] CP violation in the interference of decays with and without mixing.
This type of CP violation is possible for a subset of final states to which
both $D^0$ and $\Dzb$ can decay. 
\end{enumerate}
For a given final state $f$, CP violating contributions can be summarized 
in the parameter 
\begin{equation}
\lambda_f = \frac{q}{p} \frac{{\overline A}_f}{A_f}=
R_m e^{i(\phi+\delta)}\left| \frac{{\overline A}_f}{A_f}\right|,
\end{equation}
where $A_f$ and ${\overline A}_f$ are the amplitudes for $D^0 \to f$ and 
$\Dzb \to f$ transitions respectively and $\delta$ is the strong phase 
difference between $A_f$ and ${\overline A}_f$. Here $\phi$ represents the
convention-independent weak phase difference between the ratio of 
decay amplitudes and the mixing matrix.

The non-diagonal entries in the mixing matrix of Eq.~(\ref{MixingMatrix})
lead to mass eigenstates of neutral $D$-mesons that are different from 
the weak eigenstates,
\begin{equation} \label{definition1}
| D_{^1_2} \rangle =
p | D^0 \rangle \pm q | \bar D^0 \rangle,
\end{equation}
where the complex parameters $p$ and $q$ are obtained from diagonalizing 
the $D^0-\Dzb$ mass matrix with $|p|^2 + |q|^2 = 1$. If CP-violation
in mixing is neglected, $p$ becomes equal to $q$, so $| D_{1,2} \rangle$ 
become $CP$ eigenstates, $CP | D_{\pm} \rangle = \pm | D_{\pm} \rangle$,
\beq\label{CPeigenstates}
|D_\pm\rangle = \frac{1}{\sqrt{2}} \left[
|D^0 \rangle \pm |\barD \rangle \right]
\eeq
The mass and width splittings between these eigenstates are given by
\begin{eqnarray} \label{definition}
x \equiv \frac{m_2-m_1}{\Gamma}, ~~
y \equiv \frac{\Gamma_2 - \Gamma_1}{2 \Gamma}.
\end{eqnarray}
It is known experimentally that $\DzDzb$ mixing proceeds extremely slowly, 
which in the Standard Model is usually attributed to the absence of 
superheavy quarks destroying GIM cancellations~\cite{Golowich:2007fs,Falk:2001hx,Reviews}.
As we shall see later, this fact additionally complicates searches for CP-violation 
in charmed mesons.

\section{CP-violation in mesons}

CP violation can be searched for by a variety of methods. In general, one
can separate two ways. One way employs ``static'' observables, such as 
electric dipole moment of a baryon. Another way, more applicable to 
charm physics, employs ``dynamical'' observables, i.e. decay probabilities and
asymmetries. Here we shall concentrate on this methods of searching for 
CP-violation.

\noindent
{\it a. CP-violation in transitions, forbidden by CP-invariance.}
This method is based on the idea that if both initial and final states are
prepared as CP-eigenstates, the transition from the initial to final state 
would be forbidden if their CP-eigenvalues do not match. If CP is broken then 
transition probability would be proportional to CP-breaking parameter. 

While neither of $D$-mesons constitute a CP-eigenstates, a linear combination
of neutral $D$-mesons of Eq.~(\ref{CPeigenstates}) is. Thus such measurements
can be performed at threshold charm factories, such as CLEO-c or BES-III,
using quantum coherence of the initial state.

An example of this type of signal is a decay $(D^0 \Dzb) \to f_1 f_2$ at 
$\psi(3770)$ with $f_1$ and $f_2$ being the different final CP-eigenstates
of the same CP-parity. These types of signals are very easy to detect 
experimentally. The corresponding CP-violating decay rate for the final states
$f_1$ and $f_2$ is
\begin{eqnarray} \label{CPrate}
\Gamma_{f_1 f_2} &=&
\frac{1}{2 R_m^2} \left[
\left(2+x^2-y^2\right) \left|\lambda_{f_1}-\lambda_{f_2}\right|^2
\right.
\nonumber \\
&+& \left .\left(x^2+y^2\right)\left|1-\lambda_{f_1} \lambda_{f_2}\right|^2
\right]~\Gamma_{f_1} \Gamma_{f_2}.
\end{eqnarray}
The result of Eq.~(\ref{CPrate}) represents a slight generalization of the formula 
given in Ref.~\cite{Bigi:1986dp}. It is clear that both terms in the numerator 
of Eq.~(\ref{CPrate}) receive contributions from CP-violation of the type I 
and III, while the second term is also sensitive to CP-violation of the
type II. Moreover, for a large set of the final states the first term would be 
additionally suppressed by SU(3)$_F$ symmetry, as for instance, 
$\lambda_{\pi\pi}=\lambda_{KK}$ in the SU(3)$_F$ symmetry limit. 
This expression is of the {\it second} order in CP-violating parameters 
(it is easy to see that in the approximation where only CP violation in the mixing 
matrix is retained, $\Gamma_{f_1 f_2} \propto \left|1-R_m^2\right|^2 \propto A_m^2$).

The existing experimental constraints~\cite{HFAG} demonstrate that 
CP-violating parameters are quite small in the charm sector, regardless of 
whether they are produced by the Standard Model mechanisms or by some new physics 
contributions. Since the above measurements involve CP-violating decay {\it rates}, 
these observables are of {\it second order} in the small CP-violating 
parameters, a challenging measurement.

\noindent
{\it b. CP-violation in decay asymmetries.}

Most of the experimental techniques that are sensitive to CP violation make use of
decay asymmetries, which are similar  to the ones employed in B-physics~\cite{BigiSandaBook},
\begin{eqnarray}\label{Acp}
a_f=\frac{\Gamma(D \to f)-\Gamma({\overline D} \to {\overline f})}{
\Gamma(D \to f)+\Gamma({\overline D} \to {\overline f})}.
\end{eqnarray}
One can also introduce a related asymmetry,
\begin{eqnarray}\label{Acp1}
a_{\overline f}=\frac{\Gamma(D \to \overline f)-\Gamma({\overline D} \to f)}{
\Gamma(D \to \overline f)+\Gamma({\overline D} \to f)}.
\end{eqnarray}
For charged $D$-decays the only contribution to the asymmetry 
of Eq.~(\ref{Acp}) comes from the multi-component structure of the
$\Delta C =1$ decay amplitude of Eq.~(\ref{DirectAmpl}). In this case,
\begin{eqnarray}\label{DCPaF}
a_f &=& \frac{2 Im\left(A_1 A_2^*\right) \sin\delta}
{\left|A_1\right|^2 + \left|A_2\right|^2 + 2 Re A_1 A_2^* \cos\delta}
\nonumber \\
&=& 2 r_f \sin\phi \sin\delta,
\end{eqnarray}
where $\delta = \delta_1-\delta_2$ is the CP-conserving phase difference and 
$\phi$ is the CP-violating one. $r_f=|A_2/A_1|$ is the ratio of amplitudes.
Both $r_f$ and $\delta$ are extremely difficult to compute reliably in 
$D$-decays. However, the task can be significantly simplified if one
only concentrates on detection of New Physics in CP-violating asymmetries in the
current round of experiments~\cite{Grossman:2006jg}, i.e. at the 
${\cal O}(1\%)$ level. This is the level at which $a_f$ is currently probed
experimentally, as summarized in Table~\ref{table1}.
As follows from Eq.~(\ref{DCPaF}), in this case one should expect $r_f \sim 0.01$. 

It is easy to see that the Standard Model asymmetries are safely below this estimate.
First, Cabibbo-favored ($A_f \sim \lambda^0$) and doubly Cabibbo-suppressed 
($A_f \sim \lambda^2$) decay modes proceed via amplitudes that share the
same weak phase, so no CP-asymmetry is generated\footnote{Technically, there is 
a small,${\cal O}(\lambda^4)$ phase difference between the dominant tree $T$ amplitude
and exchange $E$ amplitudes~\cite{PetrovDPF}.}. Moreover, presence of 
NP amplitudes does not significantly change this conclusion~\cite{Bergmann:1999pm}. 
On the other hand, singly-Cabibbo-suppressed decays ($A_f \sim \lambda^1$) readily 
have two-component structure, receiving contributions from both tree and penguin 
amplitudes. In this case the same conclusion follows from the consideration of 
the charm CKM unitarity,
\beq
V_{ud} V_{cd}^* + V_{us} V_{cs}^* + V_{ub} V_{cb}^* = 0.
\eeq
In the Wolfenstein parameterization of CKM, the first two terms in this equation are of 
the order ${\cal O}(\lambda)$ (where $\lambda \simeq 0.22$), while the last one is
${\cal O}(\lambda^5)$. Thus, CP-violating asymmetry is expected to be at most 
$a_f \sim 10^{-3}$ in the Standard Model. Model-dependent estimates of this 
asymmetry exist and are consitent with this estimate~\cite{Buccella:1992sg}.
\begin{table}[t]
\begin{center}
\caption{Current experimental constraints on CP-violating asymmetries in 
charged $D$-decays~\cite{HFAG}.}
\begin{tabular}{|c|c|}
\hline \textbf{~~Decay mode~~} & \textbf{~~CP asymmetry~~} \\
\hline $D^+ \to K_S \pi^+$ & $-0.016 \pm 0.017$ \\
\hline $D^+ \to K_S K^+$ & $+0.071 \pm 0.062$ \\
\hline $D^+ \to K^+ K^- \pi^+$ & $+0.007 \pm 0.008$ \\
\hline $D^+ \to \pi^+ \pi^- \pi^+$ & $-0.017 \pm 0.042$ \\
\hline $D^+ \to K_S K^+ \pi^+ \pi^-$ & $-0.042 \pm 0.068$ \\
\hline
\end{tabular}
\label{table1}
\end{center}
\end{table}

Asymmetries of Eq.~(\ref{Acp}) can also be introduced for the neutral $D$-mesons.
In this case a much richer structure becomes available due to interplay of
CP-violating contributions to decay and mixing amplitudes~\cite{Grossman:2006jg,Bergmann:2000id}, 
\begin{eqnarray}\label{AcpNeutral}
a_f &=& a_f^d + a_f^m + a_f^i,
\nonumber \\
a_f^d &=& 2 r_f \sin\phi \sin\delta,
\\
a_f^m &=& - R_f \frac{y'}{2} \left(R_m - R_m^{-1}\right) \cos\phi,
\nonumber \\
a_f^i &=& R_f \frac{x'}{2} \left(R_m + R_m^{-1}\right) \sin\phi,
\nonumber 
\end{eqnarray}
where $ a_f^d$, $a_f^m$, and $a_f^i$ represent CP-violating contributions
from decay, mixing and interference between decay and mixing amplitudes respectively.
For the final states that are also CP-eigenstates, $f = \overline f$ and $y' = y$.

As can be seen from Eq.~(\ref{AcpNeutral}), the CP-violating asymmetries in neutral
$D$-decays depend on $\DzDzb$ mixing parameters $x'$ and $y'$. Presently, experimental 
information about the $\DzDzb$ mixing parameters $x$ and $y$ comes from the 
time-dependent analyses that can roughly be divided into two categories. 
First, more traditional studies look at the time
dependence of $D \to f$ decays, where $f$ is the final state that can be
used to tag the flavor of the decayed meson. The most popular is the
non-leptonic doubly Cabibbo suppressed decay $D^0 \to K^+ \pi^-$.
Time-dependent studies allow one to separate the DCSD from the mixing 
contribution $D^0 \to \Dzb \to K^+ \pi^-$,
\begin{eqnarray}\label{Kpi}
\Gamma[D^0 \to K^+ \pi^-]
=e^{-\Gamma t}|A_{K^-\pi^+}|^2 
\nonumber \\
~\left[
R+\sqrt{R}R_m(y'\cos\phi-x'\sin\phi)\Gamma t
\right. 
\\
\left.
+\frac{R_m^2}{4}(y^2+x^2)(\Gamma t)^2
\right],
\nonumber 
\end{eqnarray}
where $R$ is the ratio of DCS and Cabibbo favored (CF) decay rates. 
Since $x$ and $y$ are small, the best constraint comes from the linear terms 
in $t$ that are also {\it linear} in $x$ and $y$.
A direct extraction of $x$ and $y$ from Eq.~(\ref{Kpi}) is not possible due 
to unknown relative strong phase $\delta_D$ of DCS and CF 
amplitudes~\cite{Falk:1999ts}, 
as $x'=x\cos\delta_D+y\sin\delta_D$, $y'=y\cos\delta_D-x\sin\delta_D$. 
This phase can be measured independently~\cite{Gronau:2001nr}. The corresponding 
formula can also be written~\cite{Bergmann:2000id} for $\Dzb$ decay with $x' \to -x'$ and 
$R_m \to R_m^{-1}$.

Second, $D^0$ mixing can be measured by comparing the lifetimes 
extracted from the analysis of $D$ decays into the CP-even and CP-odd 
final states. This study is also sensitive to a {\it linear} function of 
$y$ via
\begin{equation}
\frac{\tau(D \to K^-\pi^+)}{\tau(D \to K^+K^-)}-1=
y \cos \phi - x \sin \phi \left[\frac{R_m^2-1}{2}\right].
\end{equation}
Time-integrated studies of the semileptonic transitions are sensitive
to the {\it quadratic} form $x^2+y^2$ and at the moment are not 
competitive with the analyses discussed above. 

Three experimental collaborations (BaBar, Belle and CDF) have recently announced
evidence for observation of $\DzDzb$ mixing~\cite{DDmixExp} using the analyses
described above. The results reported by these collaborations were combined by 
the Heavy Flavor Averaging Group (HFAG) to yield~\cite{HFAG}
\begin{eqnarray}
x &=& \left(8.4^{+3.2}_{-3.4}\right) \times 10^{-3},
\nonumber \\
y &=& \left(6.9 \pm 2.1\right) \times 10^{-3}.
\end{eqnarray}
Once again, it can be seen that the results depend on hadronic parameters, such as
the strong phase $\delta_D$. While the observed values of $x$ and $y$, which are 
believed to be dominated by the Standard Model contributions (for recent analyses of 
NP contributions see~\cite{NPinD}) and happen to be quite large, the SM CP-violating
 phases are still quite small. Thus, one can talk about almost background-free search 
for CP-violation induced by BSM interactions. Current experimental constraints on CP-violating 
asymmetries in neutral $D$-decays are summarized in Table~\ref{table2}. As one can see,
most measurements are the percent sensitivity. One should note that the rate asymmetries 
of Eq.~(\ref{Acp}) for neutral $D$-mesons require tagging of the initial state with 
the consequent reduction of the available dataset.
\begin{table}[t]
\begin{center}
\caption{Current experimental constraints on CP-violating asymmetries in 
neutral $D$-decays~\cite{HFAG}.}
\begin{tabular}{|c|c|}
\hline \textbf{~~Decay mode~~} & \textbf{~~CP asymmetry~~} \\
\hline $D^0 \to K^+ K^-$ & $+0.0136 \pm 0.012$ \\
\hline $D^0 \to K_S K_S$ & $-0.23 \pm 0.19$ \\
\hline $D^0 \to \pi^+ \pi^-$ & $+0.0127 \pm 0.0125$ \\
\hline $D^0 \to \pi^0 \pi^0$ & $+0.001 \pm 0.048$ \\
\hline $D^0 \to \pi^+ \pi^- \pi^0$ & $+0.01 \pm 0.09$ \\
\hline $D^0 \to K_S \pi^0$ & $+0.001 \pm 0.013$ \\
\hline $D^0 \to K^- \pi^+ \pi^0$ & $-0.031 \pm 0.086$ \\
\hline $D^0 \to K^+ K^- \pi^+ \pi^-$ & $-0.082 \pm 0.073$ \\
\hline
\end{tabular}
\label{table2}
\end{center}
\end{table}

One question that can be asked is what models of New Physics can be 
probed via CP-violating observables in $D$-decays in the near future. 
A decomposition of Eq.~(\ref{AcpNeutral}) allows to address this question
by studying parameters that enter Eq.~(\ref{AcpNeutral}). In particular, one
needs to study the amplitude ratio $r_f$ to see the feasibility of 
constraining a given NP model via charge asymmetries in $D$-decays.
A general conclusion of the recent study~\cite{Grossman:2006jg} is that 
${\cal O}(1\%)$ asymmetries are possible for SUSY models where new 
contributions come from QCD penguin operators and especially from 
chromomagnetic dipole operators, while tree-level direct CP violation 
in various known models is constrained to be much smaller than 
$10^{-2}$ (see Table~\ref{table3}).
\begin{table}[t]
\begin{center}
\caption{Tree-level NP contributions to $r_f$~\cite{Grossman:2006jg}.}
\begin{tabular}{|c|c|}
\hline \textbf{~~Model~~} & $\mathbf{r_f}$ \\
\hline ~Extra quarks in vector-like rep~ & $<10^{-3}$ \\
\hline ~R-parity violating SUSY~ & $<1.5\times 10^{-4}$ \\
\hline ~Two-Higgs doublet~ & $<4\times 10^{-4}$ \\
\hline
\end{tabular}
\label{table3}
\end{center}
\end{table}
Clearly, neutral D decays could exhibit contributions from indirect or direct CP 
violation (or both). One can experimentally distinguish between these 
possibilities~\cite{Grossman:2006jg} by selecting particular combinations
of final states. For instance, combined analysis of $D \to K \pi$ and 
$D \to KK$ can yield interesting constraints on CP-violating parameters, 
which are universal~\cite{Bergmann:2000id},
\bea
\Delta Y_{KK} &=& 
\frac{\Gamma'(D^0 \to K^+ K^-)-\Gamma'(\barD \to K^+ K^-)}
{\Gamma'(D^0 \to K^+ K^-)+\Gamma'(\barD \to K^+ K^-)} 
\nonumber \\
&=& a^m_{KK} + a^i_{KK},
\eea
where $\Gamma'(D^0 \to K^+ K^-)$ and $\Gamma'(\barD \to K^+ K^-)$ are the
modified decay rate parameters~\cite{Bergmann:2000id}
\bea
\Gamma'(D^0 \to K^+ K^-) &=& \Gamma_D
\left(1 \right.
\nonumber \\
 &+& \left. \eta^{CP}_f R_m(y\cos\phi-x\sin\phi)\right),
\nonumber \\
\Gamma'(\barD \to K^+ K^-) &=& \Gamma_D
\left(1 \right.
\\
&+& \left. \eta^{CP}_f R_m^{-1}(y\cos\phi+x\sin\phi)\right).
\nonumber
\eea
Here $\eta^{CP}_f=+(-)$ for CP even (odd) states. The current experimental 
world average is $\Delta Y=(-0.35\pm0.47)\times10^{-2}$, which gives a direct probe
of CP-violating asymmetries related to mixing.

\noindent
{\it c. CP-violation with untagged samples.}

It is possible to use a method that both does not require flavor or CP-tagging of the 
initial state and results in the observable that is {\it first order} 
in CP violating parameters~\cite{Petrov:2004gs}. Let's concentrate on the 
decays of $D$-mesons to final states that are common for $D^0$ and $\barD$. 
If the initial state is not tagged the quantities that one can easily measure 
are the sums 
\begin{equation} 
\Sigma_i=\Gamma_i(t)+{\overline \Gamma}_i(t) 
\end{equation}
for $i=f$ and ${\overline f}$.
A CP-odd observable which can be formed out of $\Sigma_i$ 
is the asymmetry
\begin{equation} \label{TotAsym}
A_{CP}^U (f,t) =  
\frac{\Sigma_f - \Sigma_{\overline f}}{\Sigma_f + \Sigma_{\overline f}}
\equiv \frac{N(t)}{D(t)}.
\end{equation}
We shall consider both time-dependent and time-integrated versions 
of the asymmetry (\ref{TotAsym}). Note that this asymmetry does not 
require quantum coherence of the initial state and therefore is accessible in 
any $D$-physics experiment. It is expected that the numerator and denominator 
of Eq.~(\ref{TotAsym}) would have the form,
\bea \label{NumDenum}
N(t) &=& \Sigma_f - \Sigma_{\overline f} 
= ~e^{-{\cal T}} \left[A + B {\cal T} + C {\cal T}^2 \right],
\nonumber \\
D(t) &=& ~2 e^{-{\cal T}} \left[ 
\left | A_f \right|^2 +  
\left | A_{\overline f} \right|^2 
\right],~~~~
\eea
where we neglected direct CP violation in $D(t)$.
Integrating the numerator and denominator of Eq.~(\ref{TotAsym}) over time 
yields
\beq
A_{CP}^U (f) = \frac{1}{D}\left[A + B + 2 C\right],
\eeq
where $D=\Gamma \int_0^\infty dt ~D(t)$. 

Both time-dependent and time-integrated asymmetries depend on the same parameters
$A, B$, and $C$. The result is
\bea\label{Coefficients}
A &=& \left | A_f \right|^2 -  \left | \overline A_{\overline f} \right|^2 -
\left | A_{\overline f} \right|^2 +  \left | \overline A_f \right|^2,
\nonumber \\
B &=& 
-2 y \sqrt{R} ~\left[ \sin\phi \sin \delta 
\left(
\left | \overline A_f \right|^2 + \left | A_{\overline f} \right|^2
\right) \right. \nonumber \\
&& -~ \left. \cos\phi \cos \delta
\left(
\left | \overline A_f \right|^2 - \left | A_{\overline f} \right|^2
\right)
\right],
\\
C &=& \frac{x^2}{2} A.
\nonumber
\eea
We neglect small corrections of the order of ${\cal O}(A_m x, r_f x, ...)$ and
higher. It follows that Eq.~(\ref{Coefficients}) receives contributions from both 
direct and indirect CP-violating amplitudes. Those contributions have different time
dependence and can be separated either by time-dependent analysis of Eq.~(\ref{TotAsym}) 
or by the ``designer'' choice of the final state. Note that this asymmetry is manifestly 
{\it first} order in CP-violating parameters.

In Eq.~(\ref{Coefficients}), non-zero value of the coefficient $A$ is an indication 
of direct CP violation. This term is important for singly Cabibbo suppressed (SCS) decays. 
The coefficient $B$ gives a combination of a contribution of CP violation in the 
interference of the decays with and without mixing (first term) and direct CP 
violation (second term). Those contributions can be separated by considering 
DCS decays, such as $D \to K^{(*)} \pi$ or $D \to K^{(*)} \rho$, where direct CP violation 
is not expected to enter. The coefficient $C$ represents a contribution of 
CP-violation in the decay amplitudes after mixing. It is negligibly small in the SM and 
all models of new physics constrained by the experimental data. Note that the effect of 
CP-violation in the mixing matrix on $A$, $B$, and $C$ is always subleading.

Eq.~(\ref{Coefficients}) is completely general and is true for both DCS and SCS 
transitions. Neglecting direct CP violation we obtain a much simpler expression,
\bea \label{KpiTime}
A &=& 0, \qquad C = 0, \nonumber \\
B &=& - 2 y \sin\delta \sin\phi ~\sqrt{R} ~
\left[
\left | \overline A_f \right|^2 + \left | A_{\overline f} \right|^2
\right].
\eea
For an experimentally interesting DCS decay $D^0 \to K^+ \pi^-$
this asymmetry is zero in the flavor $SU(3)_F$ symmetry limit, where
$\delta = 0$~\cite{Wolfenstein:1985ft}. Since $SU(3)_F$ is badly broken 
in $D$-decays, large values of $\sin\delta$~\cite{Falk:1999ts} are possible. 
At any rate, regardless of the theoretical estimates, this strong phase could be 
measured at CLEO-c.
It is also easy to obtain the time-integrated asymmetry for $K\pi$. Neglecting small
subleading terms of ${\cal O}(\lambda^4)$ in both numerator and 
denominator we obtain
\beq \label{KpiIntegrated}
A_{CP}^U (K\pi) = - y \sin \delta \sin \phi \sqrt{R}. 
\eeq
It is important to note that both time-dependent and time-integrated asymmetries
of Eqs.~(\ref{KpiTime}) and (\ref{KpiIntegrated}) are independent of predictions
of hadronic parameters, as both $\delta$ and $R$ are experimentally determined 
quantities and could be used for model-independent extraction of CP-violating phase 
$\phi$. Assuming $R \sim 0.4\%$ and $\delta \sim 40^o$~\cite{Falk:1999ts} and
$y \sim 1\%$ one obtains 
$\left| A_{CP}^U (K\pi) \right| \sim \left(0.04\%\right)\sin\phi$.
Thus, one possible challenge of the analysis of the asymmetry 
Eq.~(\ref{KpiIntegrated}), is that it involves a difference of two large 
rates, $\Sigma_{K^+\pi^-}$ and $\Sigma_{K^-\pi^+}$, which should be
measured with the sufficient precision to be sensitive to $A_{CP}^U$, 
a problem tackled in determinations of tagged asymmetries in $D \to K \pi$ 
transitions.

Alternatively, one can study SCS modes, where $R \sim 1$, so the 
resulting asymmetry could be ${\cal O}(1\%)\sin\phi$. However, the final states 
must be chosen such that $A_{CP}^U$ is not trivially zero. For example, 
decays of $D$ into the final states that are CP-eigenstates would result 
in zero asymmetry (as $\Gamma_f=\Gamma_{\overline f}$ for those final states) 
while decays to final states like $K^+ K^{*-}$ or $\rho^+ \pi^-$ would not.
It is also likely that this asymmetry is larger than the estimate given above 
due to contributions from direct CP-violation (see eq.~\ref{Coefficients}). 
The final state $f$ can also be a multiparticle state. In that case,
more untagged CP-violating observables could be constructed, for 
instance involving asymmetries of the Dalitz plots, such as 
the ones proposed for B-decays~\cite{Gardner:2003su}.

As any rate asymmetry, Eq.~(\ref{TotAsym}) requires either a ``symmetric'' 
production of $D^0$ and $\barD$, a condition which is automatically satisfied 
by all $p\overline p$ and $e^+ e^-$ colliders, or a correction for 
$D^0/\barD$ production asymmetry.

\section{CP-violation in baryons}

Charmed baryons provide another system for searches for CP-violation in charm.
The fact that baryons are spin-1/2 particles allows us to form CP-violating 
asymmetries that are different from the ones in the meson systems.

Taking $\Lambda_c$ as an example, a charmed baryon decay amplitude can be 
parameterized as
\beq
{\cal A}(\Lambda_c \to B \pi) = \overline{u}_B(p,s) \left[
A_S + A_P \gamma_5 \right] u_{\Lambda_c} (p_\Lambda, s_\Lambda),
\eeq
where $B$ is a charmless baryon, and $A_S$ and $A_P$ parameterize
$s-$ and $p-$wave decay amplitudes respectively. They can be combined in 
an ``asymmetry parameter'' $\alpha_{\Lambda_c}$ as
\beq
\alpha_{\Lambda_c} = \frac{2 Re\left(A_S^* A_P\right)}{\left|A_S\right|^2+\left|A_P\right|^2}.
\eeq
This parameter can be directly measured experimentally using angular distribution of
decay products in $\Lambda_c$ decay,
\beq
\frac{d W}{d \theta} = \frac{1}{2} 
\left(1 + P \alpha_{\Lambda_c} \cos\theta\right).
\eeq
Here $P$ is polarization of the initial-state baryon.
If this analysis can be done for $\overline \Lambda_c$ decay as well, then a 
CP-violating asymmetry can be formed,
\beq
{\cal A}_f = \frac{\alpha_{\Lambda_c} + \alpha_{\overline\Lambda_c}}
{\alpha_{\Lambda_c} - \alpha_{\overline\Lambda_c}},
\eeq
which follows from the fact that $\alpha_{\Lambda_c} \to - \alpha_{\overline\Lambda_c}$ 
under CP-transformation (if CP is conserved).

There were some experimental studies of this observable. In particular, 
FOCUS collaboration reported~\cite{Link:2005ft}
\beq
{\cal A}_{\Lambda\pi} = -0.07\pm 0.19\pm 0.24. 
\eeq
New studies of CP-asymmetries in charmed baryon decays are urged, which
could be performed at LHCb or even in one of the new experiments associated with 
Project-X at FNAL~\cite{Kaplan:2007aj}.

\section{Conclusions}

In summary, charm physics, and in particular studies of CP-violation,
could provide new and unique opportunities for indirect searches for New Physics. 
Large statistical samples of charm data allow unique sensitive measurements 
of charm mixing and CP-violating parameters. While unambiguous theoretical 
predictions of CP-violating asymmetries in charm transitions are hard,
observation of CP-violation at the level of ${\cal O}(1\%)$ would indicate
new physics contribution to charm decays.

\begin{acknowledgments}
This work was supported in part by the U.S.\ National Science Foundation
CAREER Award PHY--0547794, and by the U.S.\ Department of Energy under Contract
DE-FG02-96ER41005.

\end{acknowledgments}

\bigskip 


\end{document}